# A Few-Shot Learning Approach for Sound Source Distance Estimation Using Relation Networks

Amirreza Sobhdel
*Dept. of Electrical and Computer Engineering,*
*Urmia University, Iran*
E-mail: amirezasobhdel@gmail.com

Roozbeh Razavi-Far
*Faculty of Computer Science,*
*University of New Brunswick, Canada*
E-mail: roozbeh.razavi-far@unb.ca

Vasile Palade
*Centre for Computational Science,*
*Coventry University, UK*
E-mail: ab5839@coventry.ac.uk

***Abstract*—In this paper, we study the performance of few-shot learning, specifically meta learning empowered few-shot relation networks, over supervised deep learning and conventional machine learning approaches in the problem of Sound Source Distance Estimation (SSDE). In previous research on deep supervised SSDE, low accuracies have often resulted from the mismatch between the training data (from known environments) and the test data (from unknown environments). By performing comparative experiments on a sufficient amount of data, we show that the few-shot relation network outperforms other competitors including eXtreme Gradient Boosting (XGBoost), Support Vector Machine (SVM), Convolutional Neural Network (CNN), and MultiLayer Perceptron (MLP). Hence it is possible to calibrate a microphone-equipped system, with a few labeled samples of audio recorded in a particular unknown environment to adjust and generalize our classifier to the possible input data and gain higher accuracies.**

## I. Introduction

Feature extraction has been widely used to improve classification performance in many real-world applications [1]–[4]. Sound Source Distance Estimation (SSDE) is also not an exception to this, and researchers have been trying different methods to extract features from audio data to obtain better estimations. These include room impulse response method [5], direct-to-reverberant-ratio method [6], and its application in the binaural distance estimation method [7] and phase interference between observed and pseudo-observed signal waves in single-channel audio signals [8]. Among these methods, there has also been research on classical machine learning methods, which require hand-engineered data as well. Some examples for this research are the use of magnitude squared coherence as input to Gaussian Mixture Models (GMMs) for binaural audio [9] and Binaural Signal Magnitude Difference Standard Deviation (BSMD-STD) along with statistical properties of binaural sounds as input to Support Vector Machines (SVMs) and GMMs [10]. In recent years, research on SSDE and, more generally, Sound Source Localization (SSL) have leaned towards using deep learning as it has shown promising results in many applications that require extracting and processing complex features. The fed data to deep neural networks can either be hand-engineered and refined features [5], [6], [8]–[10] or data in the forms of raw wave signals or frequency domain features like spectrograms [11]–[14] directly. Deep-learning-based methods have become the main approaches for SSL, but they have two major drawbacks comparing to other approaches: (1) they require large amounts of training data, and (2) they are very sensitive to the mismatch between the training and test conditions.

In 2019, Yiwere et al. [14] trained three Convolutional Recurrent Neural Network (CRNNs) using log-scaled mel spectrograms from three rooms with different dimensions. Test results showed that the constructed models could classify the distance pretty accurately for the audio recorded in the same room that the model was trained for but for the other two rooms, the more different it was in dimensions, the less accurate the model's predictions were. These results also indicate that deep-learning-based methods for SSL and SSDE are very sensitive to the mismatch between the training and test conditions. In some cases, to address this problem, it is possible to calibrate the receiver device using a couple of example samples to give it an insight into what the input audio can be like in an unknown environment. This kind of classification is so-called few-shot learning [15].

In the recent years, research on few-shot learning has shown its power and efficiency in tackling similar problems. For example, Model-Agnostic Meta-Learning (MAML) [16], Matching Networks [17], Siamese Networks [18], Prototypical Networks [19], and Relation Networks [20] have shown promising results in few-shot image recognition. According to comparative experiments in [20], Relation Networks currently hold a state-of-the-art performance in few-shot learning. Therefore, in this work, a few-shot relation network architecture is used to address the SSDE problem, and, then, the obtained results are compared with those obtained by state-of-the-art deep learning and conventional machine learning approaches, which indicates the developed few-shot learning approach outperforms other competitors.

The rest of the manuscript is organized as follows: Section II briefly describes the problem as well as the experimental data, and pre-processing steps. Section III presents related works and Section IV presents the proposed methodology to address the SSDE problem. Section V presents the conducted experiments. Finally, the paper is concluded in Section VI.

## II. Problem Statement

Generally, we want to use a predictive model to classify the distance between the sound source and the microphone using the recorded audio as the input. But regarding the fact that every environment has a possibly different set of physical characteristics that cause a variable amount of domain and/or distribution change in the audio features, our objective is to choose a method to train a model, and, then, use it in a new environment and have comparatively optimal estimations. To simplify our experimental set-up, we will take a dataset of audio recordings from a number of different rooms, and so the goal is to better classify the distance using a model that is trained on the data from a group of different rooms.

### A. Experimental Data

In late 2016, Gaultier et al. [21] introduced a new paradigm for sound source localization referred to as Virtual Acoustic Space Traveling (VAST) and presented a first dataset designed for this purpose. VAST is a massive dataset containing simulated Room Impulse Responses (RIR) from 16 different virtual echoic rooms and one anechoic room, all different in their walls and flooring materials to some extent. The data samples are labeled with properties like source and receiver positions in the room, source and receiver absolute distance, and surface materials of the rooms. However, in this research, we merely use the echoic room RIRs and separate them based on the source and receivers' absolute distances (1, 1.5, 2, 3, and 4 meters) and room numbers.

### B. Data Pre-processing

According to [22], generally, all time-frequency representations like Mel-Spectrogram's can result in more accurate estimations and are more prosperous than baseline Mel-Frequency Cepstral Coefficient (MFCC) features, but for the reason of MFFCs being more compressed and less time-consuming during the training session, in this research, we have used MFCCs to compare the methods.

In the process of extracting these features, we make use of a window size of 1024 samples, a hop length of 256, and 33 cepstral coefficients. It is also important to mention that the audio data in VAST contains a large amount of silence, and the RIR lengths are different among 16 rooms. Therefore, before extracting MFCCs, we remove the silent parts, concatenated each sample with itself, and, then, trimmed the output if necessary to create one-second-long samples. We also change the amplitude of each sample randomly to eliminate the effect of the loudness of RIRs on the predictions.

## III. Related Works

Using an adequate number of training samples can indeed enhance predictions when classifying unseen audio samples from a new environment. Nevertheless, it is unlikely that there will be enough training data to fully generalize the model across every imaginable environment. Here, we explore potential solutions to this problem through transfer learning and few-shot learning.

Transfer learning is one of the most commonly used techniques to improve the model's performance in conditions, where there is an insufficient number of samples to train the model. Generally, transfer learning refers to the transfer of knowledge obtained from one or multiple source domains to the target domain learner to reduce the effect of insufficient amount of labelled instances in constructing the target learner [23]. There are different categorizations for transfer learning. For example, from a label-setting point of view, a transfer learning process can be one of the three categories of transductive, inductive, and unsupervised transfer learning [24]. Transductive can be used when we only have labels from the source domain, inductive can be used when we have labels from the target domain, and unsupervised can be used when we have labels from neither the source domain nor the target domain. Transfer learning approaches can also be categorized into four categories: instance-based, feature-based, parameter-based, and relational-based approaches [24].

Audio data in the form of MFCCs is complex and highly dimensional, so it is best to utilize deep neural networks for our purpose. There are different approaches for applying transfer learning in deep neural networks. The work in [25] categorizes deep transfer learning approaches into four groups: (1) Instances-based deep transfer learning: given the appropriate weights, instances in the source domain can be used in training of the target domain model; (2) Mapping-based deep transfer learning: select and map instances from source and target domains into a new data space in order to use for training the model; (3) Network-based deep transfer learning: partial reuse of weights of a model which is pre-trained on the source domain; and (4) Adversarial-based deep transfer learning: using adversarial technology to find transferable representations of features for both source and target domains.

Assuming that the target domain in our problem is the data from VAST, there are enormous audio samples that can possibly come in handy to use as a source domain, and, then, one can apply any of the mentioned deep transfer learning approaches suitable to make the final model more robust and efficient, facing audio from an unknown environment. One example of a suitable source domain dataset is Google's AudioSet dataset, a large-scale dataset of manually annotated audio events [26].

However, regardless of any amount of positive transfer of knowledge from a source domain, the final model can still suffer from the variation of the possible audio input samples from new environments. Moreover, collecting a large amount of labelled data samples from a particular environment to train is very costly and undesirable. Nevertheless, collecting one or a few samples from each class from a particular environment is possible in order to calibrate the microphone-equipped system to perform well in the environment, in which it is installed. Assessment of the performance of each of the mentioned approaches of transfer learning on the problem of SSDE would be beneficial. However, in this paper, we only resort to a variant of transfer learning called few-shot learning in order to utilize these few recorded audio samples.

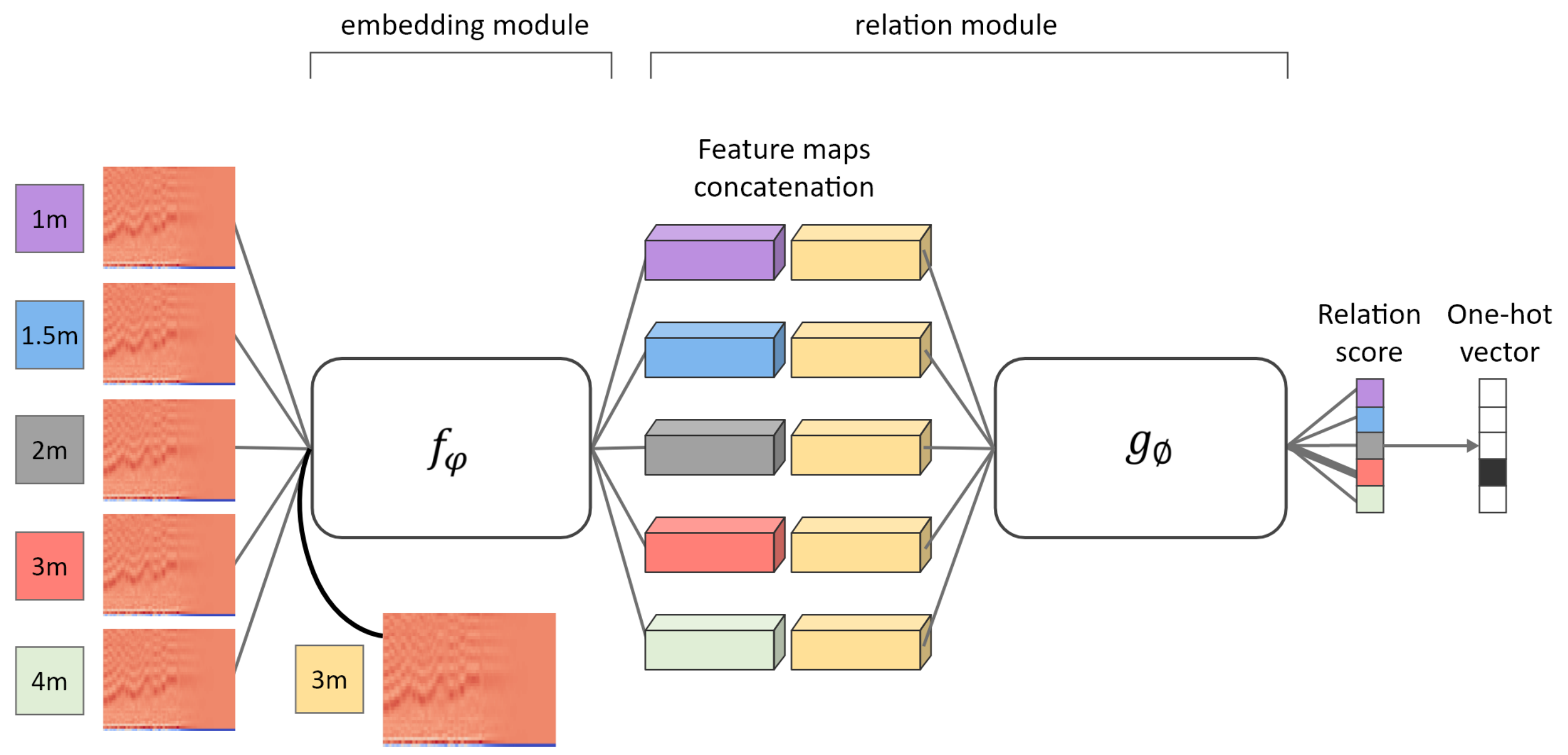

Fig. 1: The architecture of the Relation Network for a 5-way 1-shot MFCC classification task, partially adapted from [20].

**Few-shot learning:** In many real-world scenarios, which need to classify data samples, one can obtain a few samples from each existing class. This is the basis of the few-shot learning idea. According to [27], few-shot learning has a history of pre- and post-deep learning. In 2015, Koch et. al. proposed the idea of deep learning and few-shot learning together with developing the Siamese Networks [18], which aimed at learning a class-irrelevant similarity metric on pair-wise samples. This was the beginning of a new era in few-shot learning, which advanced to propose several methods for utilizing deep models along with few-shot learning, including metric learning [28], meta-learning [29], and data augmentation [30]. Specifically, meta-learning approaches have been more dominant in few-shot learning literature. Some promising works of meta-learning approaches can be Matching Networks [17], Model-Agnostic Meta-Learning (MAML) [16], Meta-Learner LSTM [31], Meta-Learning with Memory-Augmented Neural Networks (MANN) [32], Meta Networks (MetaNet) [33], Prototypical Networks [19], Relation Network [20], and Learning to Generate Matching Networks for few-shot learning (LGM-Nets) [34]. To conduct our experiments, we will use Relation Networks, which according to [20] outperformed the rest of the mentioned competitors in several benchmarks.

## IV. Methodology

In this section, we elaborate on our proposed few-shot relation method, providing a detailed overview of its components, including the embedding module, relation module, and the training process. Subsequently, we briefly introduce other competing approaches.

### A. *Few-Shot Relation Network*

In this work, a relation network identical to the relation network implemented and described in [20] is developed in order to perform our SSDE experiments. To train and test the relation network, similar to other few-shot classification models, we formally establish a training set, a support set, and a test set. The training set comprises a label space distinct from that of the support set and test set. In a target few-shot problem, if the support set contains $K$ labeled samples representative of each $C$ distinct classes, it constitutes a $C$-way $K$-shot problem.

Inspired by [17], [19], the data is fed into the network in the form of episodes, each representing a few-shot setting by having a $C$-way $K$-shot support set and a number of queries containing the same $C$ classes as the support set, both randomly selected from the training set. Generally, a relation network aims to learn a transferrable deep metric for comparing the relations between samples, which, in our case, will be represented in the form of MFCCs. Subsequently, this learned metric is utilized to output relation scores between query samples and the samples in the support set. Finally, based on these relation scores, the relation network determines which class from the support set is most similar to the query sample.

As explained in [20], a relation network consists of an embedding module $f_\varphi$ and a relation module $g_\phi$ as shown in Figure 1. In a one-shot scenario, a query sample, along with $C$ support samples each for one class, are inputted into the embedding module, which generates feature maps for these samples. The feature map of the query sample is then concatenated with each of the $C$ support sample feature maps, and the concatenated feature maps are passed through the relation module. Ultimately, the relation module produces a relation score between 0 and 1 for each combination of feature maps, indicating the similarity between the support samples and the query sample.

In a $K$-shot scenario, where $K$ is greater than 1, the output feature maps of all support samples from each class are element-wise summed to form the class's feature map. This pooled class-level feature map is then combined with the feature map of the query sample, similar to the process in the one-shot scenario. The loss function employed to train the model is Mean Squared Error (MSE), which regresses the relation scores to the ground truth. In this context, matched pairs are assigned a similarity score of 1, while mismatched pairs are assigned a similarity score of 0.

**Designed Architecture**: The embedding module of the proposed relation network consists of four convolutional blocks each containing a 64-filter 3×3 convolution, a batch normalization and a ReLU nonlinearity layer, respectively. For the first two blocks, there is also a 2×2 max-pooling layer, which follows the three mentioned layers. The relation module contains two convolutional blocks identical to the first two convolutional blocks of the embedding module (with a 2×2 max-pooling layer) and two fully connected layers with ReLU and Sigmoid functions. Figure 2 illustrates the designed architecture.

### B. *Competing Approaches*

Among the most commonly used machine learning approaches, the following approaches are selected as competitors including (1) **XGBoost**, which is one of the most effective and widely used supervised machine learning algorithms. In each of our experiments we construct an XGBoost model with 1000 boosting trees, and a max depth of 5 for each tree; (2) Support Vector Machine (**SVM**) algorithm, which is a very popular machine learning technique applied to numerous applications across various domains. Here, in each of our experiments, a SVM model is constructed along with a linear kernel; (3) **CNN** that is one of the most common and powerful supervised deep learning methods, which makes use of convolutional layers. Here, we use a standard CNN as a baseline model for the sake of comparison. Our baseline convolutional neural network consists of 3 convolution blocks and 3 fully connected layers. Each convolutional block consists of 32-filter 3×3 convolution, a ReLU nonlinearity layer, and a batch normalization layer. Following each convolutional block, there is a 2×2 max-pooling layer. Figure 3 depicts the architecture of the developed CNN; (4) **MLP** is also considered as a competing method to our proposed relation network. In this respect, a multilayer perceptron model with 4 fully connected layers has been trained for the sake of comparison.

## V. Experimental Results

To provide a comprehensive and adequate comparison of the performance between the selected state-of-the-art classic methods and the proposed few-shot relation network in the SSDE problem, we conduct experiments on various subsets of the VAST dataset. The straightforward subdivisions of the VAST consist of samples separated by room number. Furthermore, additional subdivisions can be created by combining samples from each room with samples from one or more other rooms. Initially, we trained and tested 16 CNN models using the training data from each of the 16 rooms individually. We observed high accuracies when the test set matched the room for which the model was trained, consistent with the findings reported in [14]. We also evaluated each of these 16 models on data from the other 15 rooms that the respective model had not been trained on.

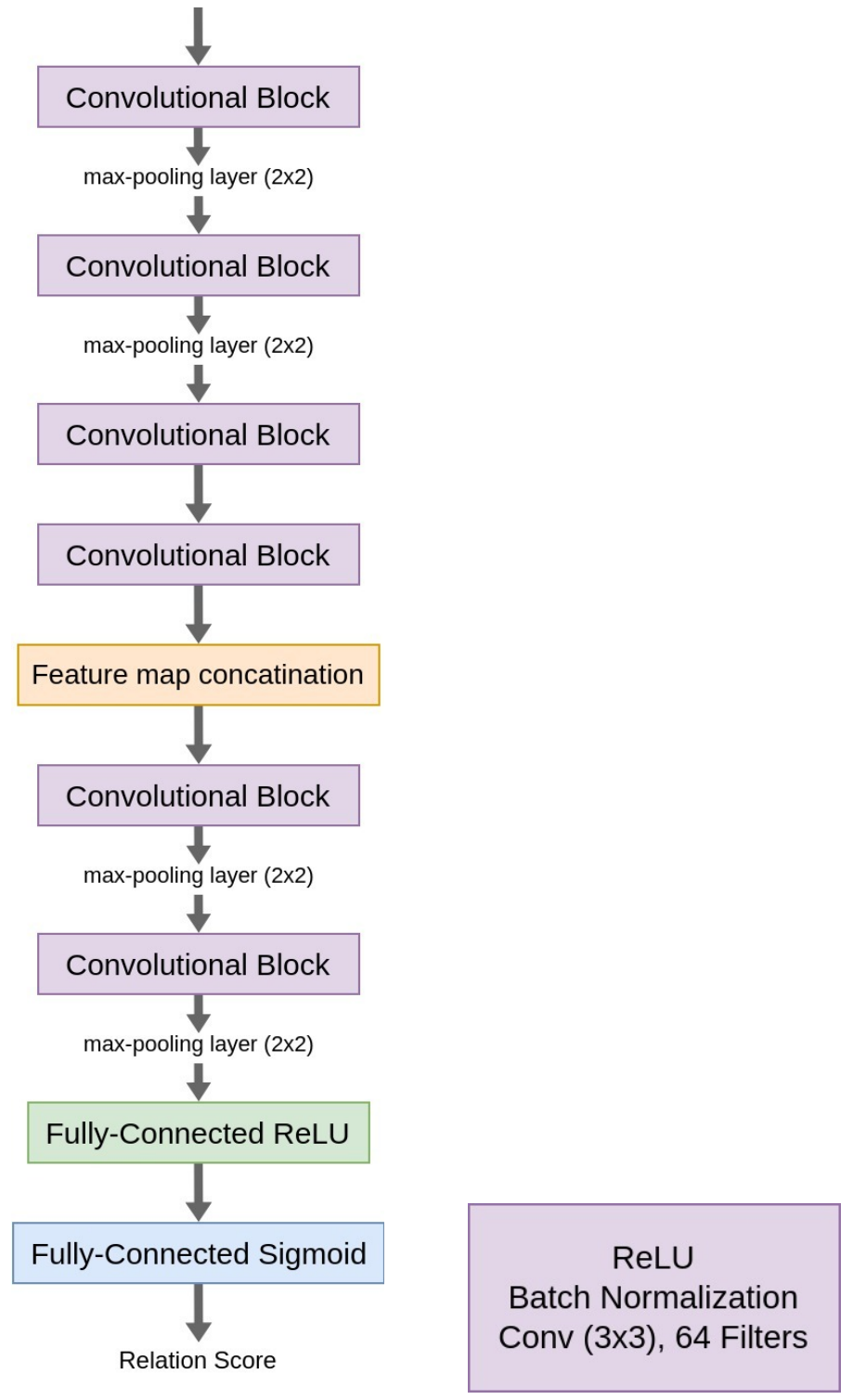

(a) Relation Network (b) Convolutional Block

Fig. 2: The architecture of the Relation Network (a) for SSDE, which contains elements including a convolutional block (b). The first four convolutional blocks (embedding module shown in Figure 1) are used to generate the feature maps for both the samples and the query, which are subsequently concatenated.

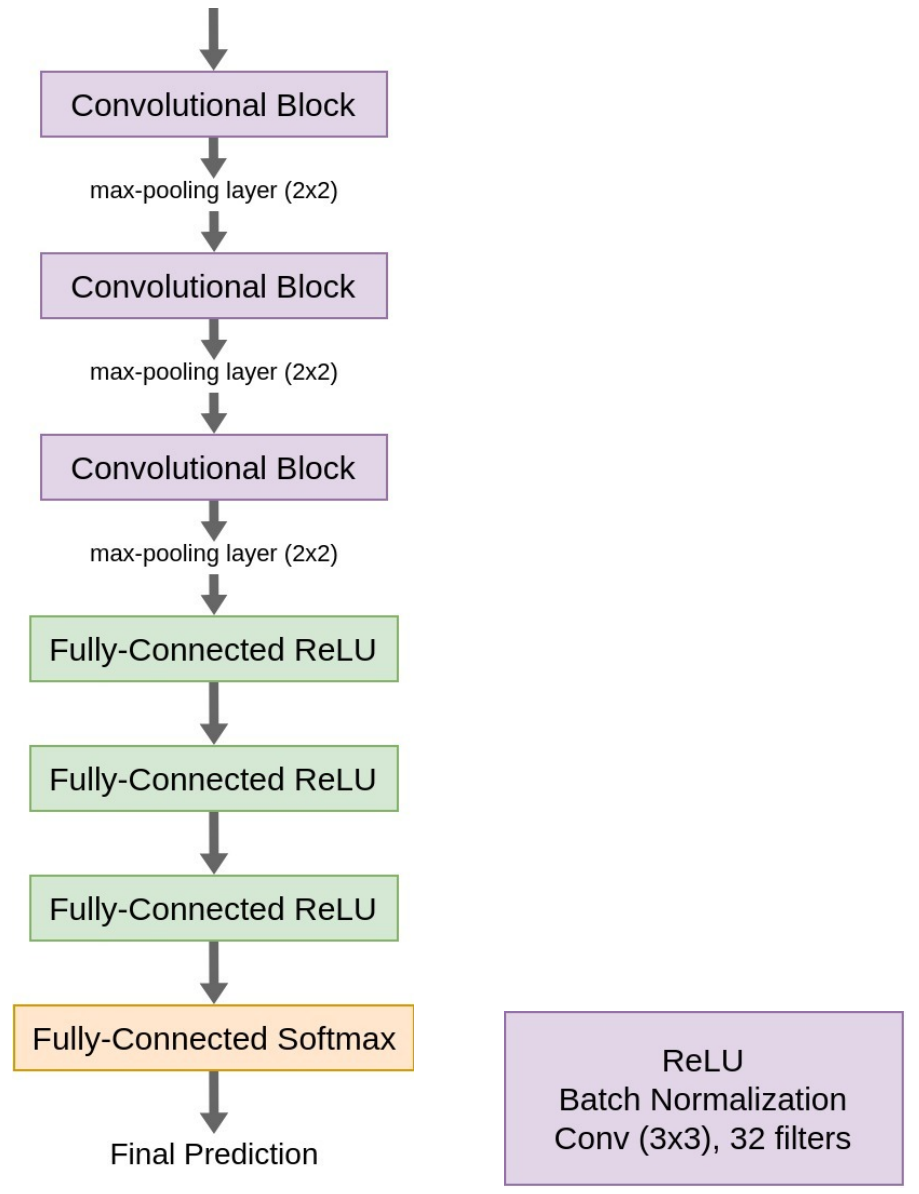

(a) Baseline CNN (b) Convolutional Block

Fig. 3: The architecture of the baseline CNN (a) for SSDE, which contains elements including a convolutional block (b).

TABLE I: Matrix representing the results of the initial CNN experiments, where cell$[i][j]$ displays the accuracy of CNN model $i$, trained using data from room number $i$, when tested on dataset $j$, which contains data from room number $j$. For instance, the accuracy of the model trained with data from room 1 (M1), tested on data from room 2 (D2), is 31.73 percent.

| | D1 | D2 | D3 | D4 | D5 | D6 | D7 | D8 | D9 | D10 | D11 | D12 | D13 | D14 | D15 | D16 |
|---|---|---|---|---|---|---|---|---|---|---|---|---|---|---|---|---|
| M1 | 99.04 | 31.73 | 31.96 | 68.94 | 98.07 | 31.77 | 31.99 | 68.93 | 36.38 | 40.39 | 40.23 | 36.80 | 37.63 | 40.76 | 40.74 | 37.26 |
| M2 | 20.49 | 99.88 | 71.18 | 12.38 | 19.49 | 98.89 | 73.20 | 11.87 | 18.29 | 44.67 | 44.24 | 15.01 | 22.28 | 44.15 | 43.21 | 19.65 |
| M3 | 24.85 | 71.91 | 99.76 | 21.67 | 23.52 | 69.84 | 96.87 | 20.76 | 25.54 | 49.78 | 44.73 | 28.36 | 17.47 | 49.56 | 44.10 | 21.78 |
| M4 | 61.54 | 32.19 | 31.63 | 98.98 | 62.86 | 32.53 | 31.77 | 96.77 | 41.19 | 40.41 | 40.23 | 48.83 | 42.34 | 40.80 | 40.74 | 45.01 |
| M5 | 95.71 | 23.65 | 31.43 | 75.78 | 97.63 | 23.51 | 32.14 | 73.12 | 36.97 | 32.71 | 41.07 | 31.71 | 35.86 | 34.16 | 40.80 | 32.39 |
| M6 | 25.00 | 99.51 | 80.14 | 15.52 | 23.90 | 99.88 | 80.66 | 15.31 | 17.22 | 48.68 | 42.15 | 33.15 | 16.79 | 49.02 | 41.73 | 31.54 |
| M7 | 23.49 | 71.10 | 96.73 | 29.37 | 23.31 | 67.15 | 99.79 | 28.50 | 48.77 | 49.45 | 41.12 | 36.27 | 41.25 | 49.71 | 41.59 | 30.16 |
| M8 | 60.94 | 32.87 | 31.46 | 95.19 | 63.90 | 32.74 | 31.85 | 98.28 | 41.31 | 41.99 | 40.95 | 47.80 | 41.52 | 42.56 | 41.57 | 39.93 |
| M9 | 32.97 | 35.98 | 35.49 | 31.95 | 33.27 | 34.43 | 35.65 | 31.91 | 97.80 | 43.71 | 38.94 | 48.67 | 90.94 | 44.45 | 39.20 | 50.40 |
| M10 | 34.78 | 34.62 | 27.37 | 28.92 | 35.27 | 34.91 | 29.18 | 28.58 | 22.20 | 98.68 | 59.96 | 41.59 | 16.29 | 96.29 | 64.30 | 40.69 |
| M11 | 21.40 | 18.40 | 22.98 | 19.49 | 21.49 | 19.56 | 24.12 | 19.22 | 19.32 | 58.23 | 99.45 | 20.52 | 22.51 | 58.93 | 98.05 | 20.03 |
| M12 | 37.30 | 31.73 | 31.66 | 44.47 | 36.95 | 31.77 | 31.80 | 44.21 | 75.89 | 41.26 | 40.27 | 97.80 | 76.93 | 41.28 | 40.78 | 91.69 |
| M13 | 30.97 | 33.06 | 24.48 | 26.83 | 30.26 | 33.51 | 23.64 | 26.44 | 88.22 | 41.24 | 39.04 | 55.40 | 98.33 | 41.66 | 40.45 | 58.81 |
| M14 | 28.74 | 32.84 | 32.03 | 23.09 | 28.62 | 32.51 | 32.12 | 22.99 | 26.65 | 96.28 | 75.41 | 13.86 | 27.88 | 98.39 | 77.75 | 13.20 |
| M15 | 27.89 | 16.08 | 37.64 | 28.14 | 26.74 | 15.37 | 40.17 | 26.87 | 12.87 | 71.22 | 97.32 | 15.68 | 8.77 | 70.87 | 99.52 | 8.40 |
| M16 | 28.17 | 31.72 | 31.66 | 29.13 | 27.97 | 31.78 | 31.79 | 32.38 | 72.07 | 40.64 | 40.27 | 89.70 | 79.59 | 40.97 | 40.80 | 98.08 |

Table I presents the results of the conducted experiments in the form of a matrix of accuracies. In this matrix, the number in cell $[i][j]$ represents the accuracy of model $i$ tested on dataset $j$, which contains samples from room number $j$. Highlighted cells correspond to cases where $i$ and $j$ are equal. It is evident that the accuracy in these cells is nearly the highest in their respective rows. Therefore, the performance of the models is highly dependent on the similarity between the test data and the data they were trained on. Upon closer examination of the results in Table I, it is apparent that the achieved accuracies in certain rooms exhibit mutual similarity to others, such as the accuracies observed for rooms 2 and 6. To provide a more comprehensive comparison of performance, we will conduct the main experiments on the VAST dataset segmented in eight different ways, as outlined in Table II. For each of the segments outlined in Table II, we train the following models including XGBoost, MLP, SVM, CNN, and Relation Network from scratch, employing random initialization and without any additional training set.

**Training the Relation Network**: Following the standard setting adopted by most existing few-shot learning works, as well as the original experiments for evaluating Relation Networks, we employed 5-way 1-shot and 5-shot configurations for the classification task. For each experiment, we trained the Relation Network with 10,000 episodes and recorded its classification accuracy by averaging the results over 500 test episodes randomly generated from the test set. In each training episode, alongside the $K$ samples, there are 15 and 10 query samples for each of the $C$ sampled classes in the 5-way 1-shot and 5-way 5-shot settings, respectively. For instance, there are $15{\times}5{+}1{\times}5 = 80$ samples in one training episode/minibatch for 5-way 1-shot experiments.

The results of these experiments, as depicted in Table III, demonstrate that few-shot relation networks surpass state-of-the-art supervised learners including CNN, MLP, SVM, and XGBoost models. This suggests that it is feasible to mitigate the impact of mismatch between training and testing data on classification accuracy by a significant margin.

## VI. Conclusion

To investigate the sound source distance estimation problem from a few-shot learning perspective, we opted to employ and adapted relation networks due to their exceptional performance simplicity, and faster execution compared to prior few-shot learning models. We conducted a comparative analysis of their performance against that of state-of-the-art supervised learners including CNN, MLP, SVM, and XGBoost. The obtained results suggest a significant improvement in addressing the conventional challenge of poor performance in unknown environments. This improvement is achieved by providing a few samples of possible input audio in that specific environment and employing a few-shot relation network. In essence, the few-shot relation network outperforms state-of-the-art supervised learners in the sound source distance estimation problem.

## References

[1] M. Farajzadeh-Zanjani, R. Razavi-Far, and M. Saif, "A critical study on the importance of feature extraction and selection for diagnosing bearing defects," in *IEEE 61st International Midwest Symposium on Circuits and Systems (MWSCAS)*, pp. 803–808, 2018.

[2] E. Hallaji, R. Aljoudi, R. Razavi-Far, M. Ahmadi, and M. Saif, *A Critical Study on the Importance of Feature Selection for Diagnosing Cyber-Attacks in Water Critical Infrastructures*, pp. 153–169. Cham: Springer International Publishing, 2021.

TABLE II: S1-S8 represent segments of data used for training and testing the models. In each of these segments, cells highlighted in green (rooms) denote test sets, comprising rooms that are mutually similar to each other. Conversely, cells shaded in gray represent training sets.

| | Room Numbers | | | | | | | | | | | | | | | |
|---|---|---|---|---|---|---|---|---|---|---|---|---|---|---|---|---|
| S1 | 1 | 2 | 3 | 4 | 5 | 6 | 7 | 8 | 9 | 10 | 11 | 12 | 13 | 14 | 15 | 16 |
| S2 | 1 | 2 | 3 | 4 | 5 | 6 | 7 | 8 | 9 | 10 | 11 | 12 | 13 | 14 | 15 | 16 |
| S3 | 1 | 2 | 3 | 4 | 5 | 6 | 7 | 8 | 9 | 10 | 11 | 12 | 13 | 14 | 15 | 16 |
| S4 | 1 | 2 | 3 | 4 | 5 | 6 | 7 | 8 | 9 | 10 | 11 | 12 | 13 | 14 | 15 | 16 |
| S5 | 1 | 2 | 3 | 4 | 5 | 6 | 7 | 8 | 9 | 10 | 11 | 12 | 13 | 14 | 15 | 16 |
| S6 | 1 | 2 | 3 | 4 | 5 | 6 | 7 | 8 | 9 | 10 | 11 | 12 | 13 | 14 | 15 | 16 |
| S7 | 1 | 2 | 3 | 4 | 5 | 6 | 7 | 8 | 9 | 10 | 11 | 12 | 13 | 14 | 15 | 16 |
| S8 | 1 | 2 | 3 | 4 | 5 | 6 | 7 | 8 | 9 | 10 | 11 | 12 | 13 | 14 | 15 | 16 |

TABLE III: Comparison of achieved accuracies by different models for each segment.

| Segment | SVM | XGBoost | MLP | CNN | Few-Shot RN |
|---|---|---|---|---|---|
| S1 | 39.40 | 61.52 | 68.65 | 73.32 | 94.21 |
| S2 | 38.60 | 50.92 | 62.05 | 68.82 | 91.00 |
| S3 | 23.14 | 61.36 | 61.03 | 69.20 | 91.37 |
| S4 | 43.87 | 66.24 | 65.00 | 71.90 | 95.11 |
| S5 | 48.52 | 53.02 | 58.91 | 65.00 | 90.02 |
| S6 | 37.44 | 52.02 | 61.35 | 64.16 | 89.00 |
| S7 | 48.14 | 50.09 | 55.05 | 61.05 | 89.45 |
| S8 | 44.03 | 55.21 | 60.64 | 67.57 | 93.50 |

[3] M. Farajzadeh-Zanjani, E. Hallaji, R. Razavi-Far, and M. Saif, "Generative-adversarial class-imbalance learning for classifying cyber-attacks and faults - a cyber-physical power system," *IEEE Trans. on Dependable and Secure Computing*, vol. 19, no. 6, pp. 4068–4081, 2022.

[4] M. Farajzadeh-Zanjani, R. Razavi-Far, M. Saif, and L. Rueda, "Efficient feature extraction of vibration signals for diagnosing bearing defects in induction motors," in *2016 International Joint Conference on Neural Networks (IJCNN)*, pp. 4504–4511, 2016.

[5] P. N. Samarasinghe, T. D. Abhayapala, M. Polettfi, and T. Betlehem, "On room impulse response between arbitrary points: An efficient parameterization," in *2014 6th International Symposium on Communications, Control and Signal Processing (ISCCSP)*, pp. 153–156, IEEE.

[6] H. Chen, T. D. Abhayapala, P. N. Samarasinghe, and W. Zhang, "Direct-to-reverberant energy ratio estimation using a first-order microphone," *IEEE/ACM Transactions on Audio, Speech, and Language Processing*, vol. 25, no. 2, pp. 226–237, 2017.

[7] Y.-C. Lu and M. Cooke, "Binaural estimation of sound source distance via the direct-to-reverberant energy ratio for static and moving sources," *IEEE Transactions on Audio, Speech, and Language Processing*, vol. 18, no. 7, pp. 1793–1805, 2010.

[8] S. Honda, N. Nakasako, T. Shinohara, T. Uebo, and M. Nakayama, "Estimation of the distance to sound source based on phase interference using cross-spectrum between actual and pseudo observed waves of a single-channel microphone," *IEEJ Transactions on Electronics, Information and Systems*, vol. 136, no. 11, p. 1525–1531, 2016.

[9] S. Vesa, "Binaural sound source distance learning in rooms," *Audio, Speech, and Language Processing, IEEE Transactions on*, vol. 17, pp. 1498 – 1507, 12 2009.

[10] E. Georganti, T. May, S. Par, and J. Mourjopoulos, "Sound source distance estimation in rooms based on statistical properties of binaural signals," *IEEE Audio, Speech, Language Process.*, vol. 21, pp. 1727–1741, 08 2013.

[11] N. Yalta, K. Nakadai, and T. Ogata, "Sound source localization using deep learning models," *Journal of Robotics and Mechatronics*, vol. 29, pp. 37–48, 02 2017.

[12] S. Adavanne, A. Politis, J. Nikunen, and T. Virtanen, "Sound event localization and detection of overlapping sources using convolutional recurrent neural networks," *IEEE Journal of Selected Topics in Signal Processing*, vol. PP, pp. 1–1, 12 2018.

[13] S. Chakrabarty and E. Habets, "Multi-speaker localization using convolutional neural network trained with noise," 12 2017.

[14] M. Yiwere and E. Rhee, "Sound source distance estimation using deep learning: An image classification approach," *Sensors*, vol. 20, no. 1, p. 172, 2019.

[15] L. Fei-Fei, R. Fergus, and P. Perona, "One-shot learning of object categories," *IEEE Transactions on Pattern Analysis and Machine Intelligence*, vol. 28, no. 4, pp. 594–611, 2006.

[16] C. Finn, P. Abbeel, and S. Levine, "Model-agnostic meta-learning for fast adaptation of deep networks," in *Proceedings of the 34th Int. Conf. on Machine Learning*, vol. 70 of *Proceedings of Machine Learning Research*, pp. 1126–1135, JMLR.org, 8 2017.

[17] O. Vinyals, C. Blundell, T. Lillicrap, K. Kavukcuoglu, and D. Wierstra, "Matching networks for one shot learning," in *Proceedings of the 30th NeurIPS*, NIPS'16, p. 3637–3645, 2016.

[18] G. Koch, R. Zemel, and R. Salakhutdinov, "Siamese neural networks for one-shot image recognition," p. 8, 2015.

[19] J. Snell, K. Swersky, and R. Zemel, "Prototypical networks for few-shot learning," in *Proceedings of the 31st Int. Conf. on Neural Information Processing Systems*, NIPS'17, p. 4080–4090, 2017.

[20] F. Sung, Y. Yang, L. Zhang, T. Xiang, P. H. Torr, and T. M. Hospedales, "Learning to compare: Relation network for few-shot learning," in *Conf. on Computer Vision and Pattern Recognition*, IEEE, June 2018.

[21] C. Gaultier, S. Kataria, and A. Deleforge, "Vast: The virtual acoustic space traveler dataset," vol. 10169, pp. 68–79, 02 2017.

[22] M. Huzaifah, "Comparison of time-frequency representations for environmental sound classification using convolutional neural networks," 2017.

[23] F. Zhuang, Z. Qi, K. Duan, D. Xi, Y. Zhu, H. Zhu, H. Xiong, and Q. He, "A comprehensive survey on transfer learning," *Proceedings of the IEEE*, vol. 109, pp. 43–76, 2019.

[24] S. Pan and Q. Yang, "A survey on transfer learning," *IEEE Trans.Knowl. Data Eng*, vol. 22, pp. 1345 1359,, Oct. 2010.

[25] C. Tan, F. Sun, T. Kong, W. Zhang, C. Yang, and C. Liu, "A survey on deep transfer learning," in *Artificial Neural Networks and Machine Learning*, pp. 270–279, Springer International Publishing, 2018.

[26] J. F. Gemmeke, D. P. W. Ellis, D. Freedman, A. Jansen, W. Lawrence, R. C. Moore, M. Plakal, and M. Ritter, "Audio Set: An ontology and human-labeled dataset for audio events," in *Proc. IEEE ICASSP 2017*, (New Orleans, LA), 2017.

[27] J. Lu, P. Gong, J. Ye, and C. Zhang, "Learning from Very Few Samples: A Survey," 9 2020. arXiv:2009.02653 [cs, stat].

[28] E. Xing, M. Jordan, S. Russell, and A. Ng, "Distance metric learning with application to clustering with side-information," in *Proc. Adv. Neural Inf. Process. Syst*, pp. 521–528, NIPS, 2003.

[29] R. Vilalta and Y. Drissi, "A perspective view and survey of metalearning," *Artificial Intell. Review*, vol. 18, no. 2, pp. 77–95,, 2002.

[30] M. Tanner and W. Wong, "The calculation of posterior distributions by data augmentation," *J. American Statistical Assoc*, vol. 82, no. 398, pp. 528–540,, 1987.

[31] S. Ravi and H. Larochelle, "Optimization as a model for few-shot learning," in *Int. Conf. on Learning Representations*, 2016.

[32] A. Santoro, S. Bartunov, M. Botvinick, D. Wierstra, and T. Lillicrap, "Meta-Learning with Memory-Augmented Neural Networks," in *Int. Conf. on Machine Learning*, pp. 1842–1850, PMLR, 2016.

[33] T. Munkhdalai and H. Yu, "Meta networks," in *Proc. Int. Conf* (M. Learn, ed.), pp. 2554–2563, 2017.

[34] H. Li, W. Dong, X. Mei, C. Ma, F. Huang, and B.-G. Hu, "Lgm-net: Learning to generate matching networks for few-shot learning," in *Proc. Int. Conf. Mach. Learn*, vol. ICML, pp. 3825–3834, 2019.